Hindawi
Complexity
Volume 2020, Article ID 1752571, 15 pages
https://doi.org/10.1155/2020/1752571*Research Article*

# Multiagent Task Allocation in Complementary Teams: A Hunter-and-Gatherer Approach

**Mehdi Dadvar** **, Saeed Moazami** **, Harley R. Myler** **, and Hassan Zargarzadeh**

*Phillip M. Drayer Electrical Engineering Department of Lamar University, Beaumont, TX 77710, USA*

Correspondence should be addressed to Hassan Zargarzadeh; hzargarzadeh@lamar.eduReceived 10 September 2019; Revised 30 December 2019; Accepted 8 February 2020; Published 26 March 2020

Academic Editor: Qingling WangCopyright © 2020 Mehdi Dadvar et al. This is an open access article distributed under the Creative Commons Attribution License, which permits unrestricted use, distribution, and reproduction in any medium, provided the original work is properly cited.Consider a dynamic task allocation problem, where tasks are unknowingly distributed over an environment. This paper considers each task comprising two sequential subtasks: detection and completion, where each subtask can only be carried out by a certain type of agent. We address this problem using a novel nature-inspired approach called "hunter and gatherer." The proposed method employs two complementary teams of agents: one agile in detecting (hunters) and another skillful in completing (gatherers) the tasks. To minimize the collective cost of task accomplishments in a distributed manner, a game-theoretic solution is introduced to couple agents from complementary teams. We utilize market-based negotiation models to develop incentive-based decision-making algorithms relying on innovative notions of "certainty and uncertainty profit margins." The simulation results demonstrate that employing two complementary teams of hunters and gatherers can effectually improve the number of tasks completed by agents compared to conventional methods, while the collective cost of accomplishments is minimized. In addition, the stability and efficacy of the proposed solutions are studied using Nash equilibrium analysis and statistical analysis, respectively. It is also numerically shown that the proposed solutions function fairly; that is, for each type of agent, the overall workload is distributed equally.## 1. Introduction

Multirobot systems are expected to undertake imperative roles in a wide variety of fields such as urban search and rescue (USAR) [1, 2], agricultural field operations [3], security patrols [4, 5], environmental monitoring [6], and industrial procedures [7]. Studies have shown that multirobot systems have advantage over single-robot systems by offering more reliability, redundancy, and time efficiency when the nature of the tasks is inherently distributed [8]. Nonetheless, the problem of multirobot task allocation (MRTA) poses many critical challenges that have called for investigation in the past two decades [9, 10]. In this regard, the complexity of MRTA problems increases significantly in a dynamic environment, where the number and location of tasks are unknown for agents [11, 12]. Thus, robots need to explore the environment to find tasks before accomplishing them. In real-world problems, any robot designated to perform one of the tasks in [1–6] needs to be sufficiently dexterous, i.e., to be equipped enough for accomplishing physical tasks such as object manipulation or rubble removal in a rescue mission [13], which inevitably make the robot relatively heavy and incapable of agile exploration. Having said that, the dynamic problem can be turned into a problem where each task is composed of sequential subtasks, each possible to be done only by a certain type of agent. In this case, for each type of subtask, a robot team of appropriate type must be employed. This case poses an unexplored MRTA problem whose coupling and cooperation between those complementary teams are the motivation of this work.

In the context of MRTA, notable attention has been devoted for revealing various aspects of dynamic problems [14–17]. For instance, Lerman et al. [18] present a mathematical model of a general dynamic task allocation mechanism where robots use only local sensing and no direct communication is needed between them for task allocation. Disregarding the communication between agents is a deficiency where the information handled by the agents plays an imperative role in the functionality of a decentralized multiagent system in a dynamic environment. In this regard, Liemhetcharat and Veloso [19] introduce a novel weighted synergy graph model to capture new interactions among agents. In the created model, agents work together in a task throughout communication where weight of the edge indicates the communication cost between agents.

In contrast to the way that Liemhetcharat and Veloso [19] utilize the communication among agents, there are works employing communications among agents to frame negotiations among them. For instance, Chapman et al. [20] pursue a decentralized game-theoretic approach in which planning is achieved via negotiation between agents. Although the results show that their approach is robust to restrictions on the agents' communication and observation range, this work is not allowing agents to have differing costs for performing the same task which makes it inapplicable to a wide variety of real-world problems. On the contrary, Michael et al. [21] propose a distributed market-based coordination algorithm in which agents are able to bid for task assignments considering each agent's cost for accomplishment of tasks to address the dynamic MRTA problem. While in real-world dynamic MRTA problems, tasks are not fully observable for all agents, the authors of this work assumed that the agents have knowledge on all tasks at a time. This assumption is too strong and does not completely reflect a dynamic environment's situation. In this regard, Sariely and Balch [22] consider a real dynamic environment, present a real-time single-item auction-based task allocation method for the multirobot exploration problem, and investigate new bid evaluation strategies.

While the works reviewed above [20–22] present different approaches to address a dynamic MRTA problem, they all similarly do not consider the agents' capabilities while developing the assignment algorithm. On this subject, Wu et al. [23] take agents' capabilities into account in order to form teams by developing a market-based novel task allocation method based on the Gini coefficient. Although the authors demonstrated that the proposed method can effectively improve the number of tasks completed by a robot system, the effect of cooperation and coupling between the formed teams is still uninvestigated. In a similar effort, Shiroma and Campos [24] model agents' capabilities as actions and utilize single-round auction to form teams and then form coordination between agents of the same team. In the same way, coupling and cooperation of the formed teams have been left unexplored in this work, though the developed framework was able to successfully resolve the required allocation issues. By the same token, Prorok et al. [25] model the multirobot system as a community of species considering agents' capabilities and then present decentralized and centralized methods to efficiently control the heterogeneous teams of robots, regardless of interaction and collaboration between those teams. Given the review above, there is a lack of critical attention paid to the cooperation and coupling between robot teams, formed based on agents' capabilities, to address a dynamic MRTA problem.

As we discussed earlier, the number and location of tasks are unknown for agents in a dynamic environment [11, 12]. In this case, robots need to explore the environment to find tasks before accomplishing them. Since tasks usually require immense efforts to be completed in real-world problems such as a rescue mission, a suitable robot needs to be equipped with various sensors and devices, much more complex mechanisms, and a higher number of actuators. As a consequence, the robots inevitably become heavy and ponderous which cannot explore the environment agilely and efficiently. Motivated by this complexity, this paper proposes a nature-inspired approach called "hunter and gatherer" which employs two teams of robots: a team of agile robots that can quickly explore an environment and detect tasks, called "hunters," and a team of dexterous robots who accomplish detected tasks called "gatherers." In fact, we are turning a dynamic MRTA problem into a problem where each task is composed of two sequential subtasks: exploration and completion. Considering this, when there are *synchronization and precedence* (SP) constraints which specify an ordering constraint for time-extended assignment (TA) problems [26], the MRTA is referred to as a TA:SP problem [27]. To the best of the authors' knowledge, the MT-MR-TA:SP problem has not been tackled in the literature so far, while it is a ubiquitous problem in a wide variety of fields such as USAR and agricultural field operations.

Consider the USAR in a disaster site in which a number of victims have got stranded in unknown locations and need immediate rescue operations. Each victim is a task that needs to be detected first and then rescued by a rescue operation that typically needs several dexterity actions, such as providing logistics supports, rubble removal, object manipulation, and in situ medical assessment and intervention [13], which make a rescue robot heavy and incapable of agile search operations. This is because a rescue robot needs to have a heavy-duty manipulator, high-power actuators, tracked locomotion mechanism, high-capacity batteries, and many sorts of sensors, cameras, and communication devices to accomplish those tasks which make the robot relatively heavy and ponderous. Hence, let us consider each task comprises two sequential subtasks: detection and completion, where each subtask can only be carried out by a certain type of robot. Thus, the case encounters an ST-MR-TA:SP or MT-MR-TA:SP problem. In the USAR example, hunters can be a group of small and light-weighted unmanned aerial vehicles (UAVs) which search the site to locate victims and gatherers can be a group of maxi-sized [13] heavy-duty unmanned ground vehicles (UGVs) that rescue detected victims relying on their dexterity capabilities.

According to the proposed hunter-and-gatherer scheme, we present a game-theoretic solution which considers coupling and cooperation between complementary agents divided into different teams by (1) utilizing market-based negotiation models, auction [28–30], and reverse auction and (2) introducing decentralized incentive-based decision-making algorithms. Proposed algorithms rely on new notions of certainty and uncertainty profit margins (CPM and UPM) that, respectively, determine the levels of confidence and conservativeness of each agent in negotiations to minimize the collective cost of task accomplishments. To enhance the effectiveness of proposed algorithms, a multitask-planning algorithm is invented for gatherer agents that enables them to queue multiple tasks in their action plan for finding the optimal solution for completing a group of tasks rather than doing one by one. We show that employing two complementary teams of hunters and gatherers can effectually improve the number of tasks completed by agents, while the collective cost of accomplishments is minimized. Moreover, the stability and efficacy of the assignment algorithms are proven by a Nash equilibrium analysis and simulation experiments, respectively. Besides, we investigated the distribution of workload, as the total costs and accomplishments of a mission, among agents and showed that the proposed algorithms function fairly; that is, for each type of agent, the overall workload is distributed equally, and all agents of the same type behave analogously under similar characteristics.

The remainder of this paper is organized as follows: The problem statement and formulation are presented in Section 2. In Section 3, the methodology including conceptual frameworks, reasoning mechanisms, and algorithms is proposed. Nash equilibrium analysis is carried out in Section 4. In Section 5, statistical analysis on simulation results is presented followed by a concluding discussion in Section 6.

## 2. Problem Statement

In this section, the problem of hunter-and-gatherer mission planning (HGMP) in the context of dynamic MRTA is explained. Assume that there are $m$ tasks distributed randomly over the environment, $E$. We consider a case that the number and the locations of tasks are unknown for agents before the execution of the HGMP. The set of tasks is denoted as $\mathcal{T} = \{T_1, \ldots, T_m\}$ in which each task is split into hunting and gathering subtasks, i.e., $T_k = \{t_k^h, t_k^g\}$ with $1 \leq k \leq m$, where $t_k^h$ and $t_k^g$ represent hunting and gathering subtasks, respectively. In this case, the set of agents is defined as $\mathcal{A} = \{A_h, A_g\}$ that comprises two teams of hunters $A_h = \{a_i^h\}$ and gatherers $A_g = \{a_j^g\}$, where $1 \leq i \leq n_h$ and $1 \leq j \leq n_g$. The cost associated with $a_i^h$ for accomplishment of $t_k^h$ is denoted as $c_{k,i}^h$, and $c_{k,j}^g$ is the cost associated with $a_j^g$ for accomplishment of $t_k^g$.

### 2.1. Assumptions. Throughout this paper, the following are assumed:

(1) Tasks are stationary; that is, they are fixed to their locations.
(2) The cost of accomplishment of each task is equal to the distance that an agent moves to do a task. An agent is considered done with a task when it reaches the task's location.
(3) All agents of the same team are identical.
(4) All agents are rational; that is, they intend to maximize their expected utility.
(5) All agents are fully autonomous and have their own utility functions; that is, no global utility function exists.
(6) Agents from complementary teams can communicate with each other using a stably connected network.

Now, the HGMP problem can be stated as follows: Suppose that there exists a tuple for the mission such that HGMP = (E, α, T). Π denotes the assignment function which assigns $m$ tasks to $n = n_h + n_g$ agents such that $\Pi: \mathcal{T} \longmapsto \mathcal{A}$. Under assumptions (1)–(6), the global objective Θ is to minimize the collective cost of Π:

$$\Theta = \min_{x_k^i, y_k^j} \left\{ \rho_h \sum_{i=1}^{n_h} \sum_{k=1}^{m} c_{k,i}^h x_k^i + \rho_g \sum_{j=1}^{n_g} \sum_{k=1}^{m} c_{k,j}^g y_k^j \right\}, \quad (1)$$

where $x_k^i$ and $y_k^j$ are binary decision variables for $t_k^h$ and $t_k^g$:

$$\begin{aligned} x_k^i \in \{1, 0\}, & \quad \forall i, k, \\ y_k^j \in \{1, 0\}, & \quad \forall j, k. \end{aligned} \quad (2)$$

In (1), weighting parameters $\rho_h$ and $\rho_g$ are introduced to sum relative collective costs of complementary teams because of the physical differences of each type Table 1.

This problem has a global objective Θ which can be achieved by determining the binary decision variables optimally. These variables need to be determined by the agents throughout explorations and negotiations in a distributed manner. Since agents are rational, each agent's objective is to maximize its own expected utility. As a consequence, the objectives of agents may be conflicting during the HGMP. Hence, the methodology should be developed so that it handles these conflicts in order to maximize the effectiveness of the HGMP and achieve the global objective Θ.

## 3. Methodology

*3.1. Conceptual Frameworks.* Hunters are assigned to explore the unknown environment. There is $I_h$ as the incentive reward for a hunter, denoted as $a_i^h$, who detects a task, denoted as $T_k$. However, the detected task can only be completed by cooperation with a gatherer. Thus, an extra incentive is added for motivating agents from complementary teams to build up a cooperation, denoted as $I_{ex}$. Hunters and gatherers involve in negotiation processes to reach agreements for completing the tasks and sharing $I_{ex}$ between themselves. In a negotiation, a hunter who has detected a task on one side and one or more gatherers on the other side are involved. An agreement determines which gatherer is assigned to complete the detected task and how much is its share from $I_{ex}$. Let us denote $0 \leq P_{k,i}^h \leq 1$ and $P_{k,j}^g = 1 - P_{k,i}^h$ as the proportions that $a_i^h$ and $a_j^g$ receive from $I_{ex}$ for accomplishment of $T_k$, respectively. Also, the gatherer who completes the detected task receives $I_g$ as a gathering incentive, when the task is completed. Since all agents are rational, they intend to maximize their incentives by accomplishing more tasks through building up more cooperation.

To establish the process by which agents come into an agreement, we define an online board on which each hunter announces the location of its new detection to find gathering partners for starting a negotiation process. Each gatherer follows the announcements on the online board to choose a waiting hunter for negotiation by analyzing the location information shared by each waiting hunter. A gatherer then sends a readiness message to the chosen hunter to start a negotiation.

Table 1: Illustration of the notations used globally in this paper.

| Symbol | Definition |
|---|---|
| $\mathcal{T}$ | The task set where $\mathcal{T} = \{T_1, ..., T_m\}$ and $1 \le k \le m$ |
| $T_k$ | The $k^{\text{th}}$ member of $\mathcal{T}$ where $T_k = \{t_k^h, t_k^g\}$ |
| $\mathcal{A}$ | The agent set where $\mathcal{A} = \{A_h, A_g\}$ |
| $a_i^h$ and $a_j^g$ | The $i^{\text{th}}$ and $j^{\text{th}}$ members of $A_h$ and $A_g$, respectively |
| $c_{k,i}^h$ and $c_{k,j}^g$ | The cost associated with $a_i^h$ for accomplishment of $t_k^h$ and the cost associated with $a_j^g$ for accomplishment of $t_k^g$, respectively |
| $n$ | Defined as $n = n_h + n_g$, where $n_h$ and $n_g$ are the total number of hunters and gatherers, respectively |
| $\Pi$ | The assignment function which assigns $m$ tasks to $n$ agents such that $\Pi: \mathcal{T} \longmapsto \mathcal{A}$ |
| $\Theta$ | The global objective which is to minimize the collective cost of $\Pi$ |
| $\rho_h$ and $\rho_g$ | The weighting parameters of hunters and gatherers, respectively |
| $x_k^i$ and $y_k^j$ | The binary decision variables for $t_k^h$ and $t_k^g$, respectively |
| $I_h$ and $I_g$ | The incentive rewards for accomplishing a hunting or gathering task, respectively |
| $I_{\text{ex}}$ | An extra incentive for motivating agents from complementary teams to build up a cooperation |
| $\vartheta$ | A generalized subscript designated to denote both hunter and gatherer agents, where: $\vartheta \in \{h, g\}$ |
| $P_{k,i}^h$ | Share of the $i^{\text{th}}$ hunter in $I_{\text{ex}}$ for hunting the $k^{\text{th}}$ task |
| $P_{k,j}^g$ | Share of the $j^{\text{th}}$ gatherer in $I_{\text{ex}}$ for gathering the $k^{\text{th}}$ task |
| $n_w^h$ | Total number of hunters that have detected a task and are waiting for starting a negotiation |
| $\alpha_h$ and $\alpha_g$ | Scaling parameters of certainty profit margins of hunters and gatherers, respectively |
| $\beta_h$ and $\beta_g$ | Scaling parameters of uncertainty profit margins of hunters and gatherers, respectively |
| $R_c^h$ and $R_c^g$ | Certainty radius for hunters and gatherers, respectively |
| $R_u^h$ and $R_u^g$ | Uncertainty radius for hunters and gatherers, respectively |
| $U_h(t_k^h)$ | The profit earned by a hunter for the $k^{\text{th}}$ task |
| $U_g(t_k^g)$ | The profit earned by a gatherer for the $k^{\text{th}}$ task |
| $\text{pi}_h$ and $\text{pi}_g$ | Profit interval of a hunter and gatherer, respectively |
| $L_{\text{low}}^h$ and $L_{\text{up}}^h$ | Lower and upper bounds of a hunter, respectively |
| $L_{\text{low}}^g$ and $L_{\text{up}}^g$ | Lower and upper bounds of a gatherer, respectively |
| $O_{i,j}^k$ | An offer made by the $j^{\text{th}}$ gatherer to the $i^{\text{th}}$ hunter for the $k^{\text{th}}$ task |
| $b_{j,i}^k$ | A bid placed by the $j^{\text{th}}$ gatherer to the $i^{\text{th}}$ hunter for the $k^{\text{th}}$ task |
| $\Upsilon_j$ | Action plan of the $j^{\text{th}}$ gatherer where $\Upsilon_j = \{v_1, ..., v_q\}$ and $q \le q_{\max}$ with $q_{\max}$ as the maximum number of tasks that a gatherer can queue |

We consider two possible scenarios in order to develop reasoning mechanisms for agents to negotiate and cooperate: (1) a waiting hunter receives only one readiness message and (2) the waiting hunter receives more than one readiness message. The first scenario resembles the bargaining or reverse auction process as there is only one buyer who aims to bargain for finding the most affordable option. The second scenario is similar to an auction process where usually there is more than one buyer interested in a specific object. We utilize these two market-based methods as negotiation models between negotiating agents. In addition, it is possible that the number of waiting hunters on the board, denoted as $n_w^h$, be more than one. In this case, the question that how a gatherer chooses a hunter among $n_w^h$ waiting agents is addressed in Section 3.5. For the time being, we assume that gatherers already know how to choose a partner and we focus on the negotiation reasoning mechanisms.

Fundamentals of reasoning mechanisms are discussed in the next section, and next, we will explain how agents rely on their reasoning mechanisms to behave in the reverse auction and auction scenarios in Sections 3.3 and 3.4, respectively.

*3.2. Reasoning Mechanism.* In this section, reasoning mechanisms for both hunters and gatherers are developed to establish their behavior during a mission that determines the way that they communicate, negotiate, and cooperate. Since fundamentals of reasoning mechanisms are similar for both type of agents, for the sake of brevity, we consider a general agent defined as $a_z^\vartheta$ with $\vartheta \in \{h, g\}$, where $1 \le z \le n_\vartheta$.

Moving on, it is time to introduce the CPM and UPM for $a_z^\vartheta$. The CPM is a circular margin with a radius of $R_c^\vartheta$, in which $a_z^\vartheta$ is certain about a profitable agreement even if its share in $I_{\text{ex}}$ is zero. The UPM is a circular margin between two concentric circles with radiuses of $R_u^\vartheta$ and $R_c^\vartheta$, in which $a_z^\vartheta$ is uncertain about making a profit in an agreement; that is, its profit strongly depends on its proportion of $I_{\text{ex}}$. Furthermore, $a_z^\vartheta$ cannot make any profit beyond its UPM even if it receives $I_{\text{ex}}$ entirely.

Figure 1 shows the CPM and UPM as two concentric circles with $a_z^\vartheta$ as the center. The agent compares its cost for accomplishing the task with its CPM and UPM to realize its state to make profitable decisions during the negotiation.

The following statements explain the states of $a_z^\vartheta$ with respect to its cost for accomplishing $t_k^\vartheta$:

  (i) State 1: if $c_{k,z}^\vartheta < R_c^\vartheta$, then $a_z^\vartheta$ can make a profit regardless of its proportion of $I_{\text{ex}}$
  (ii) State 2: if $R_c^\vartheta < c_{k,z}^\vartheta < R_u^\vartheta$, then the profit of $a_z^\vartheta$ depends on its proportion of $I_{\text{ex}}$

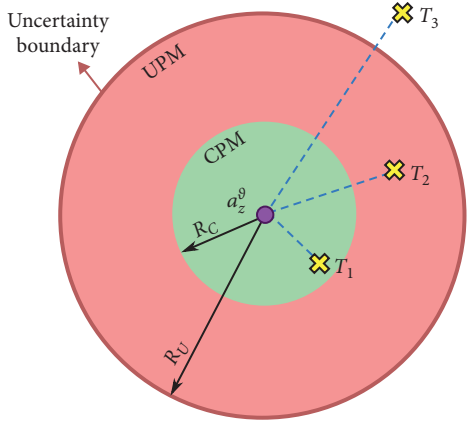

Figure 1: CPM and UPM. $T_1$ and $T_2$ are in the agent's CPM and UPM, respectively. $T_3$ is beyond the agent's uncertainty boundary.

(iii) State 3: if $c_{k,z}^\vartheta > R_u^\vartheta$, then $a_z^\vartheta$ cannot make any profit even if it receives all of $I_{ex}$

We formulate the CPM and UPM for $a_z^\vartheta$ by defining $R_c^\vartheta$ and $R_u^\vartheta$. If $a_z^\vartheta$ accomplishes $t_k^\vartheta$, then it receives $I_\vartheta$ as an incentive. Since $a_z^\vartheta$ is certain about receiving $I_\vartheta$, we have $R_c^\vartheta \propto I_\vartheta$. Hence, we define $R_c^\vartheta$ by introducing $\alpha_\vartheta$ as a scaling parameter for the CPM:

$$R_c^\vartheta = \alpha_\vartheta I_\vartheta. \quad (3)$$

In addition, $a_z^\vartheta$ receives a proportion of $I_{ex}$ that will be determined by the negotiation process, so $a_z^\vartheta$ is uncertain about its share of $I_{ex}$. Thus, we define $R_u^\vartheta$ by introducing $\beta_\vartheta$ as a scaling parameter for the UPM:

$$R_u^\vartheta = R_c^\vartheta + \beta_\vartheta I_{ex}. \quad (4)$$

Altogether, for $a_z^\vartheta$ involved in a negotiation, the utility function defined below determines its payoff.

*Definition 1* (utility function). $U_\vartheta(t_k^\vartheta)$ gives the profit earned by $a_z^\vartheta$ for accomplishing $t_k^\vartheta$ and building up a cooperation. The utility function of $a_z^\vartheta$ is defined as

$$U_\vartheta(t_k^\vartheta) = \alpha_\vartheta I_\vartheta + \beta_\vartheta \left(P_{k,z}^\vartheta I_{ex}\right) - c_{k,z}^\vartheta. \quad (5)$$

Now, we define a profit interval for $a_z^\vartheta$, regarding its state for accomplishing $t_k^\vartheta$, by which it evaluates its results in a negotiation. A profit interval is an interval for $P_{k,z}^\vartheta$ that guarantees the negotiation's profitability. According to assumption (4), $a_z^\vartheta$ wants to maximize its payoff, so in each negotiation, $a_z^\vartheta$ definitely makes a nonnegative profit such that

$$\alpha_\vartheta I_\vartheta + \beta_\vartheta P_{k,z}^\vartheta I_{ex} - c_{k,z}^\vartheta \geq 0. \quad (6)$$

This can be written as

$$\left(c_{k,z}^\vartheta - \alpha_\vartheta I_\vartheta\right)\left(\beta_\vartheta I_{ex}\right)^{-1} \leq P_{k,z}^\vartheta. \quad (7)$$

The overlap of $0 \leq P_{k,z}^\vartheta \leq 1$ and (7) yields the profit interval for $P_{k,z}^\vartheta$. The overlap in all three states is expressed as follows: If $a_z^\vartheta$ is in state 1, then $c_{k,z}^\vartheta < \alpha_\vartheta I_\vartheta$ and the left side of (7) is negative. Hence, the overlap is $0 \leq P_{k,z}^\vartheta \leq 1$; that is, in state 1, $a_z^\vartheta$ makes a profit regardless of its share in $I_{ex}$. If $a_z^\vartheta$ is in state 2, then we have $c_{k,z}^\vartheta > \alpha_\vartheta I_\vartheta$. Therefore, the overlap gives $(c_{k,z}^\vartheta - \alpha_\vartheta I_\vartheta)(I_{ex}\beta_\vartheta)^{-1} \leq P_{k,z}^\vartheta \leq 1$. And if $a_z^\vartheta$ is in state 3, then we have $c_{k,z}^\vartheta > \alpha_\vartheta I_\vartheta + \beta_\vartheta I_{ex}$, so the left side of (7) is greater than one. Hence, the overlap of (7) and $0 \leq P_{k,z}^\vartheta \leq 1$ is a null set; that is, the task is not profitable. Accordingly, the profit interval of $a_z^\vartheta$ for accomplishing $t_k^\vartheta$ is defined as $pi_\vartheta = [L_{low}^\vartheta, L_{up}^\vartheta]$, where $L_{low}^\vartheta$ and $L_{up}^\vartheta$ denote lower and upper bounds, respectively.

*3.3. The First Scenario: Reverse Auction.* Consider the scenario shown in Figure 2(a) and suppose that $a_i^h$ has detected $T_k$ at the cost of $c_{k,i}^h$, it has posted an announcement on the online board, and it receives a readiness message only from $a_j^g$. This message is a request for a quotation message; that is, $a_i^h$ offers a proportion for sharing $I_{ex}$ and $a_j^g$ decides to accept or reject the offer. Accordingly, we explain how $a_i^h$ makes offers and $a_j^g$ makes an acceptance or rejection decision, using the proposed reasoning mechanisms.

According to the process illustrated in Figure 3, $a_i^h$ calculates the lower bound, the midpoint, and the upper bound of its profit interval for making 3 offers. Since $a_i^h$ is making offers to $a_j^g$, it should send offers using $P_{k,j}^g = 1 - P_{k,i}^h$, as follows:

$$O_{i,j}^k = \left(1 - L_{low}^h\right),$$
$$O_{i,j}^{'k} = 1 - \frac{1}{2}\left(L_{up}^h - L_{low}^h\right), \quad (8)$$
$$O_{i,j}^{''k} = \left(1 - L_{up}^h\right).$$

According to Figure 3, at the first decision node, $a_i^h$ makes an offer regarding the explained process, and then at the second step, $a_j^g$ decides to accept or reject the offer. Algorithm 1 illustrates the bargaining procedure for $a_i^h$. In this algorithm, $a_{cand}^g$ denotes the gatherer agent that has sent the readiness message. In line 3 in Algorithm 1, $[O_{i,j}^k, O_{i,j}^{'k}, O_{i,j}^{''k}]$ is assigned to the vector "offers" in a random order.

Besides, $a_j^g$ uses its own profit interval to make an acceptance or rejection decision. For each received offer made by $a_i^h$, if the offer is inside $pi_g$, then $a_j^g$ accepts the offer. Otherwise, it rejects the offer.

*3.4. The Second Scenario: Auction.* Consider the auction scenario shown in Figure 2(b) and suppose that $a_i^h$ has detected $T_k$ at cost $c_{k,i}^h$, it has posted an announcement on the online board, and it receives readiness messages from $n_b \geq 2$ gatherers. In this case, $a_i^h$ holds an auction and selects the winner, where gatherers bid for sharing $I_{ex}$ to win the detected task and complete it. Accordingly, both types of agents' reasoning mechanisms need to be investigated.

We utilize the "second-price sealed-bid auction" as the negotiation framework in which the winning bidder is an agent who has placed the highest bid and it pays the amount equal to the second highest bid to the hunter holding the auction. In this auction, $a_j^g$, a gatherer bidding in the

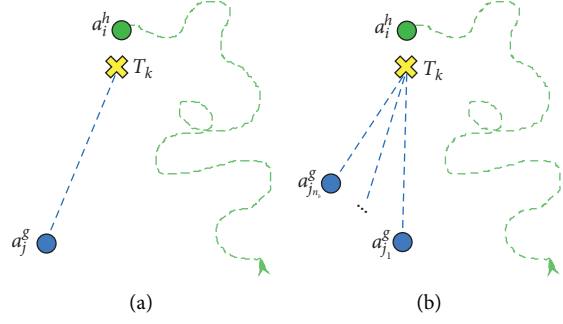

Figure 2: Possible scenarios: (a) reverse auction scenario and (b) auction scenario.

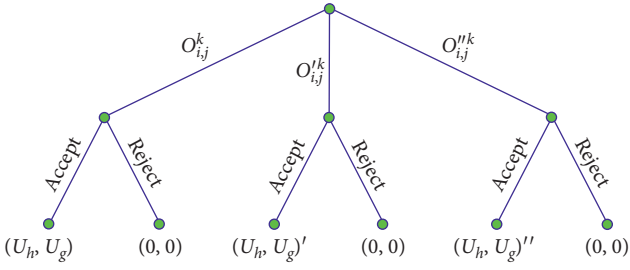

Figure 3: Reverse auction process.

```
(1)  function bargain ($c_{k,i}^h$, $\alpha_h$, $\beta_h$, $a_{cand}^g$)
(2)      $pi_h$ ⟵ calculate [$L_{low}^h, L_{up}^h$]
(3)      offers ⟵ calculate offers: [$O_{i,j}^k, O_{i,j}^{'k}, O_{i,j}^{''k}$]
(4)      for each offer do
(5)          send the offer to $a_{cand}^g$
(6)          if an acceptance message is received then
(7)              results ⟵ [$a_{cand}^g$, offer]
(8)              break for
(9)          else
(10)             results ⟵ ∅
(11)         end if
(12)     end for
(13)     return results
```

Algorithm 1: Bargaining function for a hunter agent.

auction, can bid its valuation. Since it will not pay as much as it bids if it wins, $a_j^g$ still has a chance to get a positive benefit from the auction. Therefore, the main advantage of the second-price auction over the first-price auction is that truthful bidding is an optimal strategy in a second-price auction, and as a result, it is ensured to converge to an optimal solution. Trustful bidding means that it is an optimal strategy for a bidder in a second-price auction to bid however much it values that object [31]. To that end, we explain how $a_j^g$ bids using its profit intervals first and then we discuss the way that $a_i^h$ chooses the winning bidder. $a_j^g$ bids its valuation that is the lowest bound of its profit interval. Since $a_j^g$ is making an offer to $a_i^h$ by bidding, it should send the bid using $P_{k,i}^h = 1 - P_{k,j}^g$, as follows:

$$b_{j,i}^k = \left(1 - L_{low}^g\right). \quad (9)$$

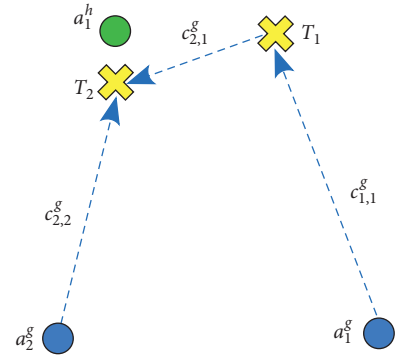

Figure 4: An example explaining the necessity of multitasking capability of gatherer agents.

Besides, $a_i^h$ chooses the winner bidder after a single round of bidding. Firstly, $a_i^h$ chooses the winning bidder, $a_w^g$, regarding the maximum bid in the set of bids, denoted as $b$, such that

$$w = \operatorname{argsmax}(b). \quad (10)$$

Secondly, $a_i^h$ checks if the second highest bid satisfies the minimum acceptable bid determined by its profit interval. Since the share of $a_i^h$ in $I_{ex}$ must satisfy (6), the minimum acceptable bid is the lower bound, $L_{low}^h$, of its profit interval such that

$$\max_{j \neq w} b_j \geq L_{low}^h. \quad (11)$$

3.5. Multitask Planning. In Figure 4, suppose that $a_1^h$ has detected $T_2$. $a_2^g$ is the only gatherer agent that can send a readiness message to start a negotiation with $a_1^h$ because $a_1^g$ is busy with gathering $T_1$. In this case, $t_2^g \longrightarrow a_2^g$ is an inefficient planning where $c_{2,1}^g < c_{2,2}^g$.

Alternatively, if $a_1^g$ is able to plan for multiple tasks at a time, it could gather $T_2$ at a lower cost. Accordingly, to prevent such ineffective plannings in the HGMP, in the following, a multitask-planning algorithm for gatherers is proposed.

Let us define an action plan in which $a_j^g$ queues multiple tasks to accomplish in the future such that $\Upsilon_j = \{v_1, \ldots, v_q\}$, where $q \leq q_{max}$ with $q_{max}$ as the maximum number of tasks that gatherers can queue. Each task has a profile in the gatherer's action plan containing required information: $v_q = \{t_k^g, c_{k,j}^g, c_{k,j}^{\prime g}, P_{k,j}^g\}$, where $c_{k,j}^{\prime g}$ denotes the temporary cost calculated by the agent for $t_k^g$. $v_1$ through $v_q$ are the tasks that $a_j^g$ is already planned to accomplish (actual tasks). In addition, $\Gamma_j$ denotes the path that $a_j^g$ follows to accomplish $\Upsilon_j$. Now, assume that $a_j^g$ wants to add a new task to $\Upsilon_j$ as $v_{i+1}$, when $q < q_{max}$. The first step is choosing a waiting hunter, and the second step is negotiating with the chosen agent. The negotiation processes have been discussed before, so here we focus on the procedure that $a_j^g$ chooses a waiting hunter to fill up $\Upsilon_j$.

The proposed method relies on the CPM and UPM to develop the gatherers' reasoning mechanism so that $a_j^g$ fills up $\Upsilon_j$ effectually. To that end, a three-step process in which $a_j^g$ chooses a waiting hunter agent for negotiation is proposed. Before starting the process, $a_j^g$ follows the online board and lists the waiting hunters in $A_{waiting}^h$ ordered by their waiting time; that is, the oldest is the first in the list. Process steps are elaborated as follows:

Step 1: $a_j^g$ considers the most prior task from $A_{waiting}^h$, denoted as $T_{cand}$. Then, $a_j^g$ plans the shortest multi-destination temporary path, using the A* search algorithm [32], denoted as $\Gamma'_j$, to gather all tasks in $\Upsilon_j$ plus $T_{cand}$. When $\Gamma'_j$ is generated, the temporary cost of each task must be updated in each task's profile.

Step 2: $a_j^g$ verifies the feasibility of making a profit from $T_{cand}$. Thus, it checks whether $T_{cand}$ is in state 3 regarding $\Gamma'_j$. If $T_{cand}$ is not in state 3, then $a_j^g$ goes to the next step. Otherwise, $a_j^g$ withdraws $T_{cand}$ and starts over from the first step.

Step 3: $a_j^g$ examines the effect of choosing $T_{cand}$ on the actual tasks in $\Upsilon_j$. When $a_j^g$ generates $\Gamma'_j$ in step 1, it may have $c_{k,j}^{\prime g} \neq c_{k,j}^g$ for the actual tasks in $\Upsilon_j$. For this reason, $a_j^g$ checks whether (6) is still true for newly calculated temporary costs for each actual task. If (6) is true for all actual tasks, $T_{cand}$ is verified for starting a negotiation process; otherwise, $a_j^g$ withdraws $T_{cand}$ and starts over from the first step.

Algorithm 2 illustrates the multitask-planning procedure for $a_j^g$. This algorithm is developed as a function for choosing a candidate task detected by a waiting hunter by considering $\Upsilon_j$. Not to mention, the output of this algorithm is not a task in the agent's action plan, i.e., $v_{q+1}$. The output is a candidate task detected by a waiting hunter that potentially can be added to $\Upsilon_j$ as $v_{q+1}$ depending on the negotiation process.

Figure 5 illustrates an example in which $a_1^g$ fills out its action plan where $q_{max} = 5$. All sequences happen before accomplishing $v_1$. In sequence 1, $a_1^g$ has already $v_1$ in its action plan and chooses $T_{C1}$ to negotiate. In sequence 2, $a_1^g$ has reached an agreement for accomplishing $T_{C1}$ and adds it to its action plan as $v_2$. Moreover, there are two candidates, and $T_{C2}$ is chosen to negotiate because $T_{C3}$ is not feasible and

```
(1)  function ChoosePartner (Υ_j, α_g, β_g, q_max,
         online_board)
(2)      A^h_waiting ⟵ list and sort waiting hunters
(3)      for each hunter in A^h_waiting α_g do
(4)          a^h_cand ⟵ the selected waiting hunter
(5)          T_cand ⟵ the task detected by a^h_cand
(6)          Γ'_j ⟵ generate a path to do Υ_j and T_cand
(7)          update Υ_j with newly calculated costs
(8)          if a_j^g is not in state 3 for T_cand then
(9)              validation_flag ⟵ true
(10)             for v_1 to v_q do
(11)                 if α_g I_g + β_g P^g_{k,j} I_ex < c'^g_{k,j} then
(12)                     validation_flag ⟵ false
(13)                 end if
(14)             end for
(15)         end if
(16)         if validation_flag == true then
(17)             results ⟵ [a^h_cand, T_cand]
(18)             break for
(19)         end if
(20)     end for
(21) return results
```

ALGORITHM 2: Multitask planning.

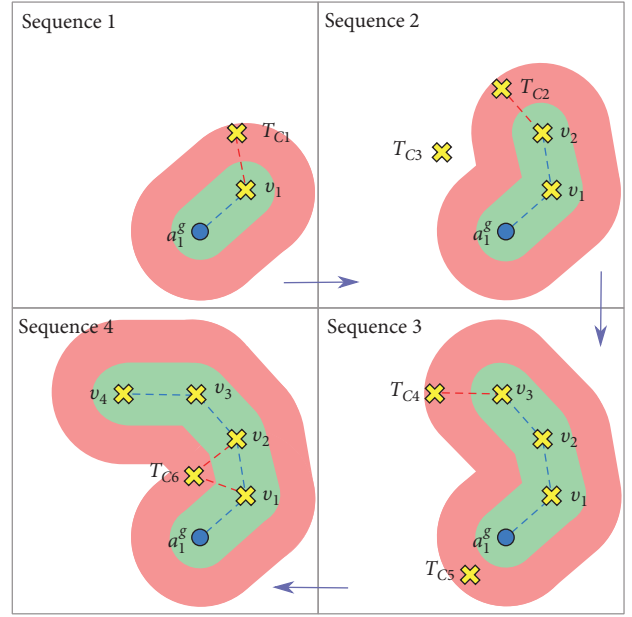

FIGURE 5: An example of multitask planning by a gatherer agent.

fails to satisfy the condition mentioned in step 2. In sequence 3, $a_1^g$ has reached an agreement for accomplishing $T_{C2}$ and adds it to its action plan as $v_3$. In addition, there are two candidate tasks, namely, $T_{C4}$ and $T_{C5}$. $T_{C5}$ cannot be verified in step 3, though it is feasible itself and passes step 2. Therefore, $a_1^g$ chooses $T_{C4}$ for negotiation. In sequence 4, $a_1^g$ has reached an agreement for accomplishing $T_{C4}$ and adds it to its action plan as $v_4$. Moreover, it chooses $T_{C6}$ to negotiate because it passes all 3 steps. Although choosing $T_{C6}$ causes change in the path to $T_2$, it does not bring $a_1^g$ into state 3 for $T_2$.

### 3.6. Decision-Making Algorithms.

Firstly, we propose a distributed decision-making algorithm determining the exploration, detection, and negotiating procedure for $a_i^h$ in the HGMP. Basically, we utilize the distributed approach to ensure the reliability of the MAS, where centralized MASs may not be robust for the reason that they are relying on a single central unit. In addition, the inherence of the proposed reasoning mechanism enables agents to make the decision independently regardless of any central unit. Furthermore, the nature of the profit margins limits all interactions between agents to local regions within the environment, so there is no need for a central unit to play a role. Even in the case of auctions, each hunter who has detected a task holds a local auction and plays the role of an action organizer temporally and locally. Having said that, a distributed decision-making algorithm fits the best to the proposed reasoning mechanisms. Accordingly, Algorithm 3 illustrates the decision-making procedure for $a_i^h$. In each iteration, $a_i^h$ explores the environment to detect a task. When $a_i^h$ detects a task, denoted as $T_{\text{detected}}$, it announces the location on the online board and waits to receive readiness messages. According to the number of readiness messages that $a_i^h$ receives, it starts a reverse auction or auction negotiation process to reach an agreement. If $a_i^h$ reaches an agreement, then it starts exploring the environment again. Otherwise, $a_i^h$ announces its detection on the online board again and does the same procedure. $\tau$ denotes the iteration number, and $\tau_{\max}$ denotes the maximum iterations in a mission.

Secondly, we present a distributed decision-making algorithm determining the negotiating and accomplishment procedure in the HGMP for $a_j^g$ regarding the explained reasoning mechanisms. Algorithm 4 illustrates the decision-making procedure for $a_j^g$. In each iteration, when $q < q_{\max}$, $a_j^g$ manages its action plan by calling the "choose partner" function first and then negotiating with the chosen hunter upon availability. If the negotiation is succeeded, then it adds the new task to $\Upsilon_j$ and updates $\Gamma_j$. Moreover, in each iteration, $a_j^g$ follows $\Gamma_j$ to gather tasks in $\Upsilon_j$. When a task is gathered, $a_j^g$ updates $\Upsilon_j$ by removing the accomplished task.

## 4. Nash Equilibrium Analysis

It is important to study the stability of the proposed algorithms to ensure that agents do not have motivation to change their behavior during the HGMP, i.e., to make sure that agents can make optimal decisions in the scenarios and do not vacillate in negotiations and task accomplishments. In this section, we study the stability of the proposed algorithms in both reverse auction and auction scenarios.

### 4.1. The First Scenario: Reverse Auction.

Consider a hunter and a gatherer agent whose preferences over outcomes are given by the utility functions $U_h(t_k^h)$ and $U_g(t_k^g)$, respectively. As shown in Figure 3, the model in which agents negotiate in the first scenario is a simplified reverse auction or bargaining process. According to assumption (6), each agent obtains sufficient information about all actions and utilities. Thus, the model turns into a perfect-information extensive form game which resembles a sharing game. We know that every (finite) perfect-information game in the extensive form has a pure-strategy Nash equilibrium (PSNE) [33]. However, the existence of PSNE does not necessarily ensure that the output of the first scenario is a PSNE. It strongly depends on the decision-making algorithm of each agent. Therefore, we need to prove if the output of the proposed reasoning mechanisms in the first scenario is a PSNE.

According to the proposed reasoning mechanisms, each agent calculates a profit interval to make the most profitable decision. To be specific, $a_i^h$ makes its best response by making

```
(1)  for τ = 1 : τ_max do
(2)      if hold == false then
(3)          explore the environment
(4)          if a new task is detected then
(5)              T_detected ⟵ the new detected task
(6)              hold ⟵ true
(7)          end if
(8)      end if
(9)      if hold == true then
(10)         announce T_detected on the online board
(11)         if one readiness message is received then
(12)             bargain
(13)         else if readiness messages >1 then
(14)             hold an auction
(15)         end if
(16)         if the negotiation is succeeded then
(17)             hold ⟵ false
(18)             mark the task as accomplished
(19)         end if
(20)     end if
(21) end for
```

ALGORITHM 3: Decision-making outer-loop algorithm of $a_i^h$.

```
(1)  for τ = 1 : τ_max do
(2)      if q < q_max then
(3)          [a_cand^h, T_cand] ⟵ choose partner
(4)          if a_cand^h ≠ ∅ then
(5)              send the readiness message to a_cand^h
(6)              if received a bargaining notification then
(7)                  bargain
(8)              else
(9)                  place a bid for the auction
(10)             end if
(11)         end if
(12)         if the negotiation is succeeded then
(13)             add T_cand to Υ_j as v_{q+1}
(14)             Γ_j ⟵ Γ'_j
(15)         end if
(16)     end if
(17)     follow Γ_j to accomplish tasks and update Υ_j
(18) end for
```

ALGORITHM 4: Decision-making outer-loop algorithm of a gatherer agent.

offers that fall into its profit interval. Similarly, $a_j^g$ makes its best response to the scenario by accepting the offers within its profit interval. In other words, the decision of each agent is its best possible response to the scenario, and it knows that the counterpart agent is also making its best response. We know that the strategy profile in which each agent is making its best response to another agent is a PSNE [33]. Consequently, the HGMP's outcome is a PSNE in the first scenario.

Although the model itself ensures the existence of PSNE and the reasoning mechanisms' outcome is a PSNE, the desirability of PSNE is still a considerable concern. The following numerical example explains the details on how scaling parameters can affect the PSNE in the first scenario.

In the reverse auction scenario, pure strategies for $a_i^h$ and $a_j^g$ are defined as $S_h = \{O_{i,j}^k, O_{i,j}^{'k}, O_{i,j}^{''k}\}$ and $S_g = \{$(AAA), (AAR), (ARA), (ARR), (RAA), (RAR), (R, R, A), (RRR)$\}$, respectively, where A and R stand for acceptance and rejection actions of $a_j^g$, respectively. Now, let us assume that $I_h = I_g = I_{ex} = 10$, $\alpha_h = \beta_h = 1$, and $\alpha_g = \beta_g = 1$. If we have $c_{k,h}^h = 8$ and $c_{k,j}^g = 10$ for accomplishing $T_k$, then offers are calculated from (8), as follows: $O_{i,j}^k = 0$, $O_{i,j}^{'k} = 5$, and $O_{i,j}^{''k} = 10$. Hence, {(A, A, A), offer1} is one of the equilibria; that is, $a_j^g$ accepts the first offer which results $U_g(t_k^g) < 0$. On the contrary, if we only change scaling parameters of $a_j^g$ such that $\alpha_g = \beta_g = 0.9$, then {(A, A, A), offer1} is no longer a PSNE. Instead, {(R, R, R), offer1} is a PSNE; that is, $a_j^g$ rejects the first offer. In conclusion, the desirability of PSNE in the first scenario can be guaranteed by designating appropriate scaling parameters $\alpha_h$, $\beta_h$, $\alpha_g$, and $\beta_g$.

*4.2. The Second Scenario: Auction.* In the second scenario, we investigate the existence of NE by a theorem based on the CPM and UPM concepts. We investigate 3 conditions to find the NE in an auction process. We will prove the theorem by contradiction; that is, we show that no agent, involving in an auction scenario, has a motivation to deviate from a strategy profile which satisfies all 3 conditions.

**Theorem 1.** *Consider the HGMP in the second scenario associated with the second-price sealed-bid auction with participation of a hunter and gatherers whose preferences over outcome are given by the utility functions $U_h(t_k^h)$ and $U_g(t_k^g)$, respectively. Then, b is a Nash equilibrium if and only if conditions (i) and (ii) are satisfied for $w = \arg\max(b)$ and condition (iii) is satisfied for the hunter agent:*

(i) $\max_{j \neq w}(\alpha_g I_g + \beta_g I_{ex} - c_{k,j}^g) \leq b_w$, *i.e., the winner submitted a sufficiently high bid*

(ii) $\max_{j \neq w} b_j \leq \alpha_g I_g + \beta_g I_{ex} - c_{k,w}^g$, *i.e., the winner's valuation is sufficiently high*

(iii) $\alpha_h I_h + \beta_h(P_{k,i}^h I_{ex}) - c_{k,i}^h \geq 0$, *i.e., the second highest bid satisfies the minimum bid determined by the hunter*

*Proof.* If (i) does not hold, $\max_{j \neq w}(\alpha_g I_g + \beta_g I_{ex} - c_{k,j}^g) > b_w$, then $a_{j \neq w}^g$ has an interval to increase its bid, $[k_{low}^g, P_{k,w}^g]$, in which it can lower its share to $k_{low}^g$ and place even a higher bid than $b_w$ and win the auction. Hence, $a_{j \neq w}^g$ has a motivation to deviate and increase its payoff. If (ii) does not hold, $\max_{j \neq w} b_j > \alpha_g I_g + \beta_g I_{ex} - c_{k,w}^g$, then for $a_w^g$, denoted as the winner, we have $U_g(t_k^g) < 0$; that is, its payoff is negative. Therefore, it can deviate by submitting a losing bid and increasing its payoff to 0. Finally, if (iii) does not hold, $\alpha_h I_h + \beta_h(P_{k,i}^h I_{ex}) - c_{k,i}^h < 0$, then the hunter agent's payoff is negative for the second highest bid. Thus, it can deviate by rejecting all bids and increase its payoff to 0 because it has a strong motivation to hold another auction in the following iterations and avoid a negative payoff. □

Nevertheless, the existence of NE does not necessarily ensure that the scenario's output is a NE. It strongly depends on the decision-making algorithm of each agent. In this regard, we know that each gatherer involving in the auction scenario places a bid according to (9). This means each gatherer agent bids its own valuation, i.e., $b_j = v_j$. Accordingly, conditions (i) and (ii) are always true because not only the winner has placed the highest bid among all bidders but also it does not have a negative payoff. Besides, the hunter agent is using (10) to choose the winning bidder and (11) to verify the minimum requirement satisfaction of the second highest bid. Hence, condition (iii) is also true. As a conclusion, according to Theorem 1 and also the decision-making algorithms of all agents participating in an auction, the result of the auction scenario is a NE.

## 5. Simulation Results

In this section, we present simulation results to (1) validate the fairness of the proposed algorithms, i.e., to ensure that the overall workload is distributed equally among agents of both types, by comparing agents' effectiveness in a set of experiments and analyzing the results by paired *T*-test and ANOVA [34] methods, (2) study the effect of profit margins on the total effectiveness of the HGMP, (3) demonstrate the efficacy of the proposed multitask-planning algorithm for gatherers by investigating its effect on the HGMP's total effectiveness, and (4) verify the functionality of the hunter-and-gatherer scheme, i.e., considering each task comprising two sequential exploration and completion subtasks, by a comparison between the HGMP and a basic alternative method in which each agent does both hunting and gathering tasks itself.

To simulate the proposed approaches, we developed a multirobot simulation platform in MATLAB from scratch. In this platform, we can implement the simulations on any custom map, while the number of agents of each type is adjustable. We provide some basic functions for each type of agent to enable them maneuver over the determined environment. For gatherers, we utilized the $A^*$-based motion planning algorithm which enables them to move along two points in a grid environment. In addition, we provided a basic frontier-based exploration algorithm [35] for hunters. Besides, the number of tasks is also adjustable while they get located randomly over the environment. As a matter of fact, we also provided the perpetual mode for implantation of the simulations where for each gathered task, another task will be distributed randomly in the environment. Accordingly, at each iteration, there are a certain number of tasks available in the environment which is adjustable for each mission. Furthermore, in the perpetual mode, each explored and known grid of the environment turns into an unknown grid after certain iterations. The perpetual mode helps the analysis be done in a much more accurate and evidence-based way.

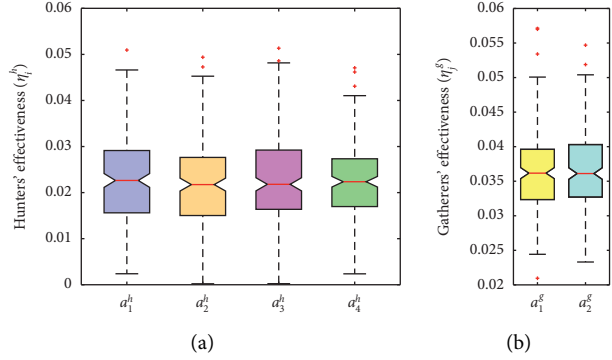

Figure 6: Investigating the fairness of the HGMP for (a) hunter agents and (b) gatherer agents. Statistically, for each type of agent, there is no significant difference in agents' effectiveness that demonstrates the fairness of the HGMP.

All simulations have been executed under the following conditions: (1) the environment is sectioned as an $e_L \times e_W$ grid of tiles where $e_L = e_W = 100$, (2) the quantities of each type of agent are adjusted as $n_h = 4$ and $n_g = 2$, (3) there are always $m_p = 50$ tasks in the environment, (4) the maximum number of iterations is determined as $\tau_{\max} = 1000$, (5) the rewards are assigned to be $I_h = I_g = I_{ex} = 140$, and (6) we considered the weighting parameters as $\rho_h/\rho_g = 0.2$.

### 5.1. Fairness of the HGMP.
To demonstrate that the accomplishment workload is distributed equally for each type of agent, the concept of fairness is introduced. To that end, we define an effectiveness factor for each agent of both types based on their costs and accomplishment. Then, using the statistical analysis, we prove the fairness of the HGMP by comparing effectiveness of different agents of each type. Let $\eta_i^h$ and $\gamma_i^h$ denote the effectiveness of $a_i^h$ and the number of tasks hunted by the agent, respectively, as follows:

$$\eta_i^h = \frac{\gamma_i^h}{\sum_{k=1}^{m} c_{k,i}^h x_k^i}. \qquad (12)$$

Similarly, $\eta_j^g$ and $\gamma_j^g$ denote the effectiveness of $a_j^g$ and the total number of tasks gathered by the agent, respectively, such that

$$\eta_j^g = \frac{\gamma_j^g}{\sum_{k=1}^{m} c_{k,j}^g y_k^j}. \qquad (13)$$

Figure 6(a) shows the statistical results of $\eta_i^h$ for all hunters in 200 missions. As $n_h > 2$, an ANOVA test has been applied to the collected data to statistically prove the fairness of the HGMP for hunters. The ANOVA test has been applied as follows: $H_0$: $\mu_1^h = \mu_2^h = \mu_3^h = \mu_4^h$, $H_1$: $\mu_i^h \neq \mu_i^h$, and $\alpha = 0.05$, where $\mu_i^h$ denotes the average of $\eta_i^h$ for $a_i^h$ in 200 tests and $\alpha$ denotes the significance level. According to the results of the ANOVA test, $F = 0.377$, $F - \text{crit} = 2.61$, and $p$ value $= 0.77$. Since $F < F - \text{crit}$ and $p$ value $> \alpha$, we have to retain the null hypothesis. Thus, it has been proved that $\mu_1^h = \mu_2^h = \mu_3^h = \mu_4^h$, which means that there is no significant difference between averages of hunters' effectiveness in 200 tests.

In addition, as $n_g = 2$, a paired $T$-test has been applied to the data to investigate the fairness of the HGMP for gatherers. The hypothesis testing has been done in a manner such that $H_0$: $\mu_1^g - \mu_2^g = D_0$, $H_1$: $\mu_1^g - \mu_2^g \neq D_0$, $D_0 = 0$, $n_s = 200$, dof $= 199$, and $\alpha = 0.05$. According to the test, $p$ value $= 0.315$. Since $p$ value $> \alpha$, we must retain the null hypothesis. Therefore, it has been proven that $\mu_1^g - \mu_2^g = D_0 = 0$, as it is illustrated in Figure 6(b), which means that there is no significant difference between averages of gatherers' effectiveness in 200 tests.

Both statistical analyses indicate that all agents of the same type behave analogously under similar characteristics. In fact, this analysis numerically validates the Nash equilibrium analysis proved for the HGMP. It means that if the fairness concept investigated above is not valid for the HGMP and favors certain agents unfairly, then there are strong motivations for other agents to deviate from the proposed negotiation structure.

### 5.2. Effects of Agents' Profit Margins on Mission's Effectiveness.
The effects of scaling parameters of profit margins, $\alpha_g$, $\beta_g$, $\alpha_h$, and $\beta_h$, on the total effectiveness of the HGMP need to be investigated in order to show the functionality of the CPM and UPM for both types of agents. To that end, we define an effectiveness factor for the HGMP, $\eta_t$, which is the ratio of the total number of completed tasks, $\gamma_t$, and the collective cost of the whole mission, $C_t$, as follows:

$$C_t = \rho_h \sum_{i=1}^{n_h} \sum_{k=1}^{m} c_{k,i}^h x_k^i + \rho_g \sum_{j=1}^{n_g} \sum_{k=1}^{m} c_{k,j}^g y_k^j, \qquad (14)$$

$$\gamma_t = \sum_{j=1}^{n_g} \gamma_j^g, \qquad (15)$$

$$\eta_t = \frac{\gamma_T}{C_T}. \qquad (16)$$

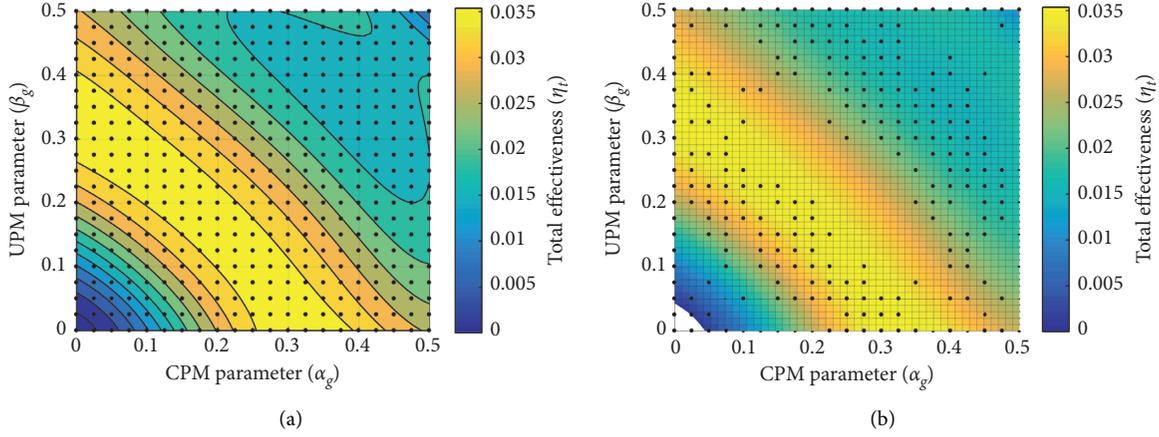

Figure 7: Investigating the effect of gatherers' CPM and UPM on the HGMP's effectiveness $\eta_t$: (a) contour plot of the results using a polynomial curve fitting and (b) polynomial curve fitting.

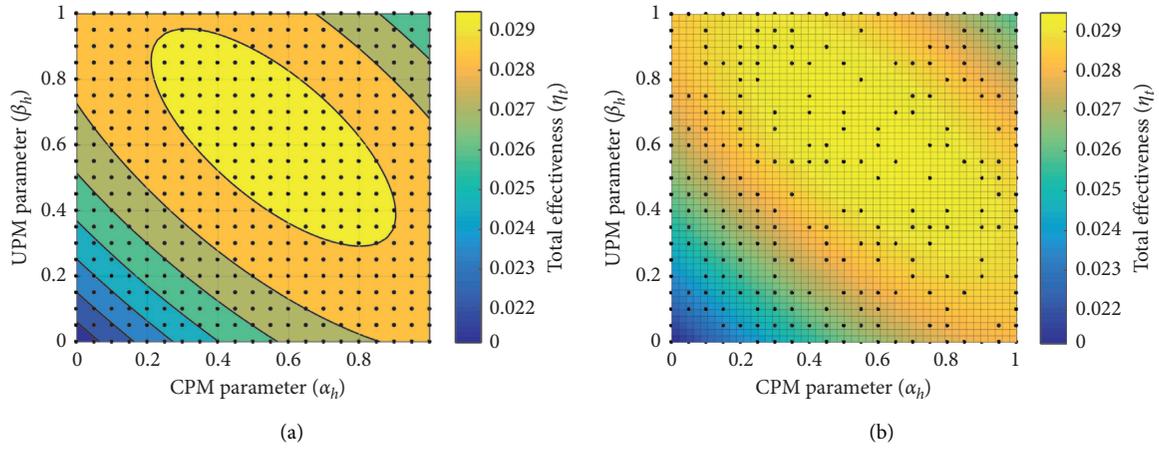

Figure 8: Effect of hunters' CPM and UPM on the HGMP's effectiveness $\eta_t$: (a) contour plot of the results using a polynomial curve fitting and (b) polynomial curve fitting.

We ran the algorithms for all values of $\alpha_g$ and $\beta_g$ that are multiples of 0.025 such that $0 \leq \alpha_g \leq 0.5$ and $0 \leq \beta_g \leq 0.5$, while $\alpha_h = \beta_h = 0.35$. $\eta_t$ has been calculated for each set of values for $\alpha_g$ and $\beta_g$, as illustrated in Figure 7. Basically, this figure explains the correlation between the total effectiveness of the HGMP and the profit margin parameters of gatherers. The yellow area shows the area in which the total effectiveness is maximum. In this figure, the horizontal and vertical axes are $\alpha_g$ and $\beta_g$, respectively, i.e., the scaling parameters of gatherers, and the color mapping represents the total effectiveness of the HGMP, i.e., $\eta_t$. According to the results, $\eta_t$ is vanishingly small when $\alpha_g + \beta_g < 0.2$ which means agents cannot reach an agreement for completing the detected tasks. Furthermore, $\eta_t$ reaches its maximum when $0.2 < \alpha_g + \beta_g < 0.5$. Next, $\eta_t$ falls gradually when $\alpha_g + \beta_g > 0.5$ because each gatherer's CPM and UPM are large so that the agent does not fall into state 3 and easily reach any agreement. As a result, each gatherer accomplishes a significant number of tasks inefficiently which reduces $\eta_t$.

In the same way, we ran the algorithms for all values of $\alpha_h$ and $\beta_h$ that are multiples of 0.05 such that $0 \leq \alpha_h \leq 1$ and $0 \leq \beta_h \leq 1$, while $\alpha_g = \beta_g = 0.15$. $\eta_t$ has been calculated for each set of values for $\alpha_h$ and $\beta_h$, as illustrated in Figure 8. Basically, this figure explains the correlation between the total effectiveness of the HGMP and the profit margin parameters of hunters. The yellow area shows the area in which the total effectiveness is maximum. In this figure, the horizontal and vertical axes are $\alpha_h$ and $\beta_h$, respectively, i.e., the scaling parameters of hunters, and the color mapping represents the total effectiveness of the HGMP, i.e., $\eta_t$. Accordingly, $\eta_t$ is too low when $\alpha_h + \beta_g < 0.4$ approximately, which means the CPM and UPM of hunters are too small and only a few agreements are reached. Then, for $\alpha_h + \beta_g > 0.4$, $\eta_t$ increases gradually to reach its maximum and then again decreases.

According to the proposed reasoning mechanism, when the scaling parameter of an agent's CPM decreases, the agent gets less confident. And when the scaling parameter of an agent's UPM increases, the agent gets less conservative. In this regard, for both types of agents, the best strategy to reach the maximum of $\eta_t$ is neither being completely confident nor being fully conservative, but a combination of both leads to the optimum result.

The oblique yellow area in Figure 7(b), exposing the maximum values of $\eta_t$, is much narrower than the one in Figure 8(b). It shows that the CPMs and UPMs of gatherers have a more distinct influence on $\eta_t$ than the ones of hunters. The rationale behind this dissimilarity is that hunters rely on their CPMs and UPMs after hunting a task, i.e., after accomplishing a task, and then consider them only for finding a gatherer to complete the task. On the contrary, gatherers consider their CPMs and UPMs before gathering a task, i.e., before any accomplishment. Consequently, this difference causes a much more distinct influence of gatherers' CPMs and UPMs on $\eta_t$.

### 5.3. The Effect of Multitask Planning on the HGMP's Effectiveness.

In this section, we aim to study the effect of the proposed multitask-planning algorithm for gatherers on the total effectiveness of the HGMP defined in (16). Accordingly, we investigate the effect of $q_{max}$, which is the queue size of each gatherer, on $\eta_t$. To that end, we ran 200 missions for each value of $q_{max}$, varying from 1 to 10, and measured $\eta_t$ in each mission, as illustrated in Figure 9.

To understand how much $\eta_t$ increase when $q_{max}$ changes from $q_{max} = 1$ to $q_{max} = 10$, we applied a paired $T$-test to the two of collected data sets. The first data set contains 200 measures of $\eta_t$ for $q_{max} = 1$, and the second data set comprises 200 measures of $\eta_t$ for $q_{max} = 10$. The test has been conducted considering $H_0: \mu_2 - \mu_1 \leq D_0$, $H_1: \mu_2 - \mu_1 > D_0$, $D_0 = 0.7\mu_1$, $n_s = 200$, $dof = 199$, and $\alpha = 0.05$, where $\mu_1$ and $\mu_2$ denote the average of $\eta_t$ for the first and second data sets, respectively. According to the test result, $p$ value = 0.0004, $t = 3.39$, and $t_{0.05,199} = 1.65$. Since $t > t_{0.05,199}$ and $p$ value $< \alpha$, we reject $H_0$. Therefore, the results prove that $\eta_t$ increases more than 70% by changing $q_{max}$ from 1 to 10. Moreover, the results also show that the HGMP remains fair for gatherer agents by increasing $q_{max}$. Figure 10 demonstrates that there is no significant difference between effectiveness of two gatherers for each value of $q_{max}$.

Besides, Figure 11 shows how the HGMP's total effectiveness converges for different values of $q_{max}$ in a manner such that $q_{max} = 1$, $q_{max} = 4$, and $q_{max} = 10$. According to the results, by increasing the value of $q_{max}$, $\eta_t$ becomes more variant and the convergence time decreases, while $\eta_t$ enhances significantly as was proven before.

### 5.4. Functionality Validation of the HGMP by a Comparison.

In this section, we are intended to analyze the functionality of the proposed hunter-and-gatherer scheme. As discussed before, we consider a dynamic problem to be a TA:SP problem where each task is composed of two sequential detection and completion subtasks. Although we have discussed different aspects of the proposed approach in the previous sections, here we want to explicitly compare the proposed approach with an alternative approach in which there is only one type of agent doing both exploration and completion of tasks together.

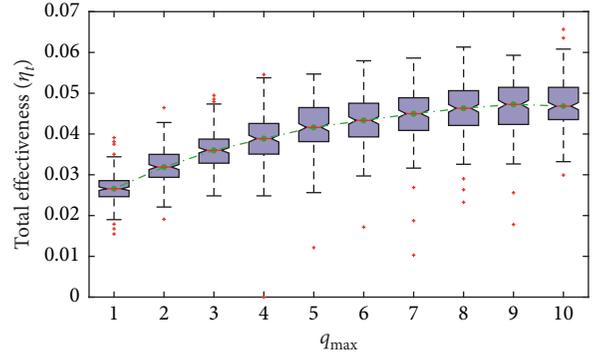

Figure 9: Investigating the effect of $q_{max}$ on $\eta_t$. The results show a significant increase in $\eta_t$ when we increase $q_{max}$ from 1 to 10.

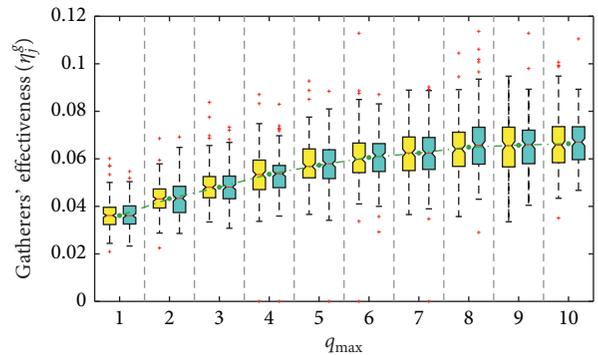

Figure 10: Fairness of the HGMP for gatherer agents for each value of $q_{max}$. The results demonstrate that the workload is still distributed equally on both gatherers when $q_{max}$ increases.

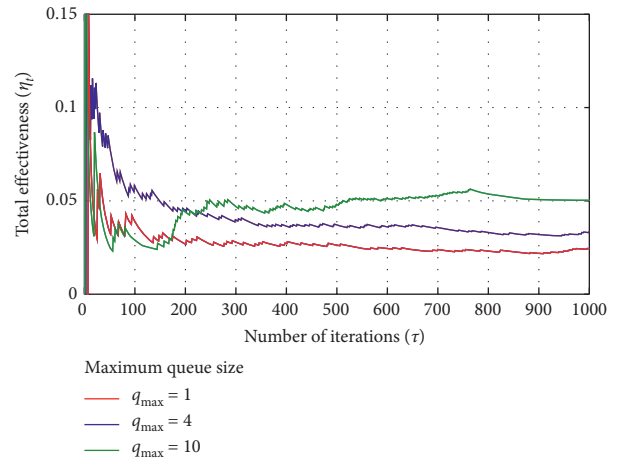

Figure 11: Convergence of the HGMP's total effectiveness for $q_{max} = 1$, $q_{max} = 4$, and $q_{max} = 10$, during $\tau_{max} = 1000$ iterations.

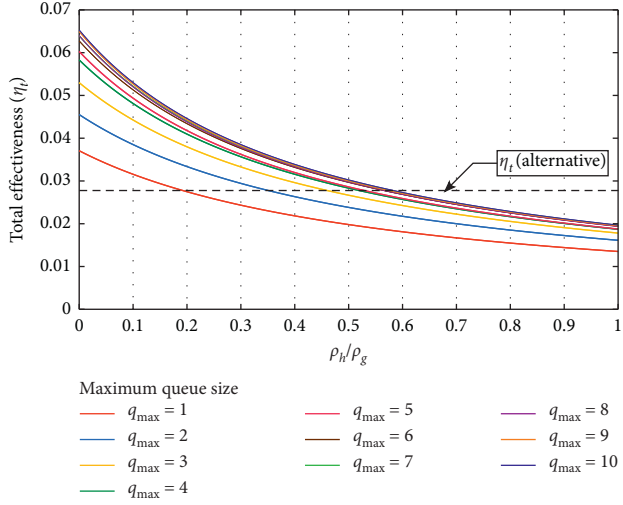

Figure 12: Comparing the HGMP with an alternative model where each agent does both exploration and completion together.

According to the rationale behind the hunter-and-gatherer approach, hunters must be more agile and cost-efficient in exploration and maneuvering. Therefore, we first plotted the total effectiveness of the HGMP with respect to $\rho_h/\rho_g$ which ranges from $\rho_h \ll \rho_g$, i.e., $\rho_h/\rho_g = 0$, to $\rho_h \approx \rho_g$, i.e., $\rho_h/\rho_g = 1$. Second, we ran the explained alternative approach to be able to judge the HGMP's functionality. Since in this approach there is no hunter-and-gatherer scheme, we only have one type of agent and the obtained total effectiveness is dependent on the ratio $\rho_h/\rho_g$. By this comparison, we basically wanted to answer the following question: Is the HGMP profitable compared to the alternative method? Figure 12 shows the results of the implemented simulations for that purpose, as explained above. Thus, the answer is that it depends to the ratio $\rho_h/\rho_g$ and this is why we ended up in having a criterion for the HGMP to be profitable. According to the results, for $\rho_h/\rho_g < 0.2$, the HGMP has distinct advantage in terms of $\eta_t$ over the alternative model for any value of $q_{max}$. Furthermore, the HGMP still remains advantageous for $\rho_h/\rho_g < 0.6$. Consequently, it is economic to employ the HGMP for the stated dynamic problem if and only if we utilize the hunter-and-gatherer agent that satisfies $\rho_h/\rho_g < 0.6$. In other words, if we employ two robots from different types as a hunter and gatherer such that the hunter's cost for following a certain path is less than 0.6 of the gatherer's cost for following the same path, then employing the HGMP will be profitable. Considering the USAR example, the hunter can be a small UAV, while the gatherer should necessarily be a heavy-duty UGV. If we consider the cost as the power consumption, then the $\rho_h/\rho_g < 0.6$ criterion will be satisfied easily.

A screen capture video of the simulation results can be found as a supplementary material along with this paper, by using the YouTube link "youtu.be/HJuiP5DMZfo," or by scanning the following QR code.

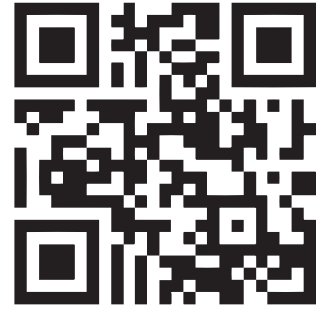

## 6. Conclusion

Inspired by the problem of "MRTA in an unknown environment," we proposed the idea of task allocation based on coupling and cooperation between complementary teams in a hunter-and-gatherer scheme. Furthermore, this work presented distributed reasoning mechanisms relying on the notions of certainty and uncertainty profit margins in which levels of confidence and conservativeness are modeled, while an effective multitask-planning algorithm for gatherers is proposed that allows them to queue multiple tasks for finding the optimal solution for completing a group of tasks rather than doing one by one. By comparing the proposed hunter-and-gatherer scheme with an alternative method, where there is only one type of agent doing both exploration and completion of tasks together, we established a criterion to judge profitability of the proposed method. Examining the real-world problems mentioned earlier confirms that the profitability criterion is reasonably satisfiable. We also found that the extreme behavior of an agent, being too confident or too conservative, hurts the total effectiveness of the mission. Furthermore, statistical analysis demonstrates a significant improvement of total effectiveness effected by the multitask-planning algorithm. However, while computational complexities for execution of the multitask-planning algorithm manifold by increasing the size of an agent's queue size, the total effectiveness of the HGMP does not increase linearly.

Future works will consider the problem of adjusting the scaling parameters by an agent during a mission to achieve the optimal performance from both agent and team points of view. We also intend to develop a multirobot exploration algorithm based on the notions of profit margins in the context of dynamic MRTA problems and investigate the effect of different multirobot exploration algorithms on the HGMP.

## Data Availability

The data and source code used to support the findings of this study are available from the corresponding author upon request.

## Conflicts of Interest

The authors declare that they have no conflicts of interest.

## Acknowledgments

This research was supported by the Lamar University via internal grants.